\documentclass[aip,jcp,superscriptaddress,reprint,showpacs,amsfonts,amsmath]{revtex4-1}
\usepackage[final]{graphicx}
\usepackage{comment}
\usepackage[normalem]{ulem}

\let\bs\boldsymbol
\DeclareMathOperator\tr{tr}
\newcommand{\tlname}[1]{\ensuremath{\text{\textit{#1}}}}

\usepackage[usenames,dvipsnames]{xcolor}
\usepackage[normalem]{ulem}

\begin{document}
\title{Channel Flow of a Tensorial Shear-Thinning Maxwell Model:
  Lattice Boltzmann Simulations}
\date\today
\newcommand\dlr{\affiliation{Institut f\"ur Materialphysik im Weltraum,
  Deutsches Zentrum f\"ur Luft- und Raumfahrt (DLR), 51170 K\"oln,
  Germany}}
\newcommand\ukn{\affiliation{Zukunftskolleg and
  Fachbereich Physik, Universit\"at Konstanz,
  78457 Konstanz, Germany}}
\newcommand\uoe{\affiliation{SUPA, School of Physics and Astronomy,
  University of Edinburgh, Mayfield Road, Edinburgh, EH9~3JZ, Scotland}}
\author{S.~Papenkort}\ukn
\author{Th.~Voigtmann}\ukn\dlr

\begin{abstract}
We introduce a nonlinear generalized tensorial Maxwell-type constitutive
equation to describe shear-thinning glass-forming fluids, motivated by a recent
microscopic approach to the nonlinear rheology of colloidal suspensions.
The model captures a nonvanishing dynamical yield stress at the glass
transition and incorporates normal-stress differences.
A modified lattice-Boltzmann (LB) simulation scheme is presented that
includes non-Newtonian contributions to the stress tensor and deals with
flow-induced pressure differences. We test this scheme in pressure-driven
2D Poiseuille flow of the nonlinear generalized Maxwell fluid. In the
steady state, comparison with an analytical solution shows good agreement.
The transient dynamics after startup and cessation of the pressure
gradient are studied; the simulation reproduces a finite stopping time
for the cessation flow of the yield-stress fluid in agreement with previous
analytical estimates.
\end{abstract}

\maketitle

\section{Introduction}

Flow problems invovling complex fluids are ubiquituous in nature and
industry \cite{Larson,Oswald}.
They often probe the nonlinar-response regime, as for example in
shear-thinning fluids, where the effective viscosity decreases rapidly
with increasing shear rate and hence depends sensitively on the flow
geometry. Since such flows are no longer described by the standard
Newtonian linear-response behavior where shear stress and shear rate
are linearly related, they are termed non-Newtonian.

Ultimately, one would like to understand the microscopic mechanisms of
non-Newtonian fluids and how these determine the macroscopic flow properties.
This is a formidable task, usually approached in two steps: one first tries
to construct a constitutive equation that acts as a closure relation to the
governing Navier-Stokes (or generally, continuum mechanics) equations at the
mesoscopic level.
In a second step, these classical field-theory equations are solved, usually
numerically.
Most commonly, constitutive equations are ad-hoc assumptions guided
by experimental data \cite{Bird}. One of the few exceptions is the
(linear) rheology of polymer melts,
based on the seminal work of Doi and Edwards \cite{DoiEdwards}.

In the case of dense glass-forming fluids, shear thinning prevails as
the dominant nonlinear mechanisms for not too strong strain rates.
Progress towards deriving
constitutive equations starting from the microscopic equations of motion
has been possible because the slow relaxation of density fluctuations
proceeds by generic mechanisms where the precise details of the microscopic
interactions, beyond excluded-volume entropic forces, are less important.
Specifically for colloidal suspensions, starting from the many-body
advection-diffusion equation, an integration-through-transients (ITT)
formalism combined with mode-coupling theory (MCT)
\cite{Fuchs.2002,Brader.2007,Brader.2008} has been successful. This approach
is in particular aimed at describing the interplay between slow dynamics close
to the glass transition and externally imposed flow. In principle,
this route yields a constitutive equation that is fully determined by the
microscopic interactions of the system. However, to date, it is still not
possible to treat this even numerically. Even schematically simplified
ITT-MCT models \cite{Brader.2009} are difficult to solve beyond the
steady state \cite{Voigtmann.2012}.

To make some progress, a simplified model has been proposed that incorporates
some of the essential ideas and findings of ITT-MCT. Starting from the
Maxwell model of linear visco-elastic fluids, one incorporates the
acceleration of the slow structural-relaxation dynamics by shear
\cite{FuchsCatesFaraday,FuchsBallauffCSA}.
This nonlinear generalized Maxwell (nlM) model has been used in systems under
homogeneous simple shear to discuss, e.g.,
qualitative features of the flow curves close to the glass transition
\cite{Voigtmann.2011} or aspects of creeping flow \cite{Siebenbuerger.2012}
and large-amplitude oscillatory shear \cite{Voigtmann.2013pre}.
Below we present the extension of this model to arbitrary incompressible
flow and incorporate it into the low-Mach-number Navier-Stokes equations
to address the pressure-driven Poiseuille flow through channels.

Idealized glass formers are yield-stress fluids, i.e., at the glass transition
they are characterized by a nonvanishing stress even in the limit of small
strain rates. The signature of a yield stress in Poiseuille flow is the
appearance of a non-parabolic velocity profile that is almost flat in the
center of the channel, causing the fluid to move as a ``plug'' in this
inner region where the applied force is not able to overcome the yield
stress. This has been confirmed in direct molecular-dynamics simulations
of a glass-forming fluid \cite{Varnik.2008}. We confirm below that the
simple nlM model captures this pheonomenology and allows for an analytical
solution in steady state. The transient evolution from equilibrium to
steady state after application of the driving pressure, and the relaxation
back to the quiescent state are studied by numerically solving the
corresponding Navier-Stokes equations. A particular feature of the glass
forming fluid that is captured naturally by the nlM model is the appearance of
large normal-stress differences that cause the central plug to be subject
to an additional pressure. In channels with constant cross-section this
does not influence the laminar flow profiles, but flow-induced pressure
changes may play a decisive role in understanding shear-localization
phenomena that are characteristic of many amorphous materials
\cite{Besseling.2010}.

One mesoscopic method to solve the Navier-Stokes equations for small
velocities is the Lattice Boltzmann (LB) simulation
\cite{Succi,Duenweg.2009}. This method has
become increasingly popular over the last decades, because of its
conceptual simplicity and because it is well suited for parallel computing.
On a suitably chosen spatial lattice, one introduces
local densities corresponding to discrete velocity vectors. The LB
simulation evolves these densities by a sequence of streaming and
collision steps. The collision operator relaxes the local distribution
towards an equilibrium form, suitably chosen to ensure that in the
continuum limit one recovers the Navier-Stokes equations for a Newtonian
fluid. The simplest form, the Bhatnagar-Gross-Krook (BGK) collision operator,
involves just one relaxation time $\tau_\text{LB}$. This is then directly
related to the Newtonian shear viscosity.

A simple way of extending the LB method to deal with
non-Newtonian fluids, pioneered by Aharonov and Rothman \cite{Aharonov.1993},
is to adjust the BGK relaxation time $\tau_\text{LB}(\dot\gamma)$ locally
depending on a (scalar) measure of the local shear rate $\dot\gamma$.
However, this adjusts just one scalar property, retaining the
Newtonian-fluid-like relation between shear and bulk viscosities and
neglecting tensorial aspects of the constitutive equation. One should
then rather speak of a generalized Newtonian fluid with a locally varying
effective viscosity $\eta_\text{eff}(\dot\gamma)$.
Nevertheless, the steady-state results have been shown to be rather accurate
in simple flow geometries, whenever analytical results are available for
comparison \cite{Gabbanelli.2005,Boyd.2006}.
In recent years, a number of studies have employed variants of
this generalized-Newtonian
LB, addressing also more complicated geometries and the flow through
porous media, see e.g.\ Refs.~\onlinecite{Kehrwald.2005,Ouared.2005,
Boyd.2007,Yan.2008,Ashrafizaadeh.2009,Wang.2009,Tang.2010}, or the extension
to an implicit scheme regarding the calculation of the collision rate
\cite{Vikhansky.2008}.
(A recent more comprehensive overview of a growing body of work is
found in Ref.~\onlinecite{Phillips.2011}.)
The generalization to an LB scheme on unstructured grids proposed by
Succi and coworkers also uses this description of
a generalized Newtonian fluid \cite{Pontrelli.2009}.
The temporal evolution of the local relaxation time can be implemented by a
finite-difference
scheme in order to capture time-dependent thixotropic rheology, leading
to a class of hybrid-LB algorithms
\cite{Derksen.2009,Derksen.2009b}.

To incorporate truly non-Newtonian fluids into LB, addressing their tensorial
character, one can adjust the LB equilibrium distribution function
in a stuiable manner. Such algorithms are closely related to so-called
lattice kinetic schemes \cite{Junk.1999,Junk.1999b,Inamuro.2002}.
One may also exploit that the gradient of the non-Newtonian
stresses appears equivalently to an external force density in the
Navier-Stokes equations.
The dynamics of the non-Newtonian forces is then traced either by a modified
LB scheme at the cost of introducing an enlarged set of lattice-node
densities \cite{Denniston.2001,Sulaiman.2006,Malaspinas.2010}, or
through suitable finite-difference solvers in hybrid-LB schemes
\cite{Marenduzzo.2007b,Henrich.2010,Frantziskonis.2011}.

In this paper, we present a modified LB scheme that allows to naturally
incorporate non-Newtonian stresses, including flow-induced pressure and
normal-stress differences relevant close to the glass transition.
While our scheme can be extended to a hybrid-LB method, we focus here on
a constitutive equation that is local in time for simplicity.
The method is outlined in Sec.~\ref{lattice-boltzmann}.
We then describe the constitutive equation based on the nonlinear generalized
Maxwell model in Sec.~\ref{maxwell-model}, together
with some analytical results used to check the accuracy of the simulation.
We then present the results for pressure-driven 2D channel flow
in Sec.~\ref{results}.

\section{Lattice Boltzmann Simulations}\label{lattice-boltzmann}

\subsection{Method for Non-Newtonian Fluids}

We consider a fluid of local mass density $\rho(\vec r,t)$ and velocity
$\vec u(\vec r,t)$, whose evolution is described by the Navier-Stokes
equations for the momentum density $\vec\jmath(\vec r,t)=\rho(\vec r,t)
\vec u(\vec r,t)$,
\begin{gather}
  \label{continuityrho}
  \partial_t\rho+\vec\partial\cdot\vec\jmath=0\,,\\
  \label{ns}
  \partial_tj_\alpha+\partial_\beta(\rho u_\alpha u_\beta)
    =-\partial_\alpha p_0+\partial_\beta\Sigma_{\alpha\beta}+f^\text{ex}_\alpha
  \,.
\end{gather}
Greek indices denote the Cartesian components of the fields, and summation
over repeated indices is implied. In the momentum flux, there appears
the Eulerian advection term $\rho\vec u\vec u$, possibly an externally imposed
body-force density $\vec f^\text{ex}$, and a surface term
encoded in the stress tensor $\boldsymbol\Pi=p_0\boldsymbol1-\boldsymbol\Sigma$.
The thermodynamic pressure $p_0$ is assumed to depend on the conserved
quantities only; it is given by an equation of state.
Non-uniform flow gives rise to stress terms that depend on the velocity
gradients, $\boldsymbol\Sigma=-\delta p\boldsymbol1+\bar{\boldsymbol\Sigma}$;
these are split into an isotropic ``nonequilibrium pressure'' $\delta p
=-(1/d)\tr\boldsymbol\Sigma$, where $d$ is the spatial dimension,
and a traceless deviatoric part $\bar{\boldsymbol\Sigma}$. (We use an
overbar to denote the traceless part of any tensor.) These quantities
need to be supplied by a rheological constitutive equation.

Newtonian fluids obey the simplest admissible linear constitutive equation,
$\boldsymbol\Sigma^\text{N}=\eta\bar{\boldsymbol D}+\eta_b
\boldsymbol1\tr\boldsymbol\kappa$, where $\boldsymbol D=\boldsymbol\kappa
+\boldsymbol\kappa^T$, and an overbar denotes the traceless part of a
tensor. $\kappa_{\alpha\beta}=\partial_\beta u_\alpha$
is the velocity-gradient tensor.
In the following, it will be useful to separate a Newtonian contribution
from the overall stress tensor, identifying the non-Newtonian stresses,
$\boldsymbol\Sigma=\boldsymbol\Sigma^\text{N}+\boldsymbol\Sigma^\text{nN}$.

For incompressible flow, $\vec\partial\cdot\vec u=0$, and the equation of
state can be reduced to its first-order density variation,
$p_0(\rho)\approx p_0(\rho_0)+c_s^2(\rho-\rho_0)$, where $\rho_0$ is some
reference density, and $c_s$ is the speed of sound.
The lattice Boltzmann simulation approximates the Navier-Stokes equations
for incompressible flow at low Mach numbers, $\tlname{Ma}\sim u/c_s\ll1$,
by allowing for weak compressibility.
In treating non-Newtonian fluids, care has to be taken since the
nonequilibrium pressure may be nonzero even in incompressible flow. Due
to the equation of state, this translates to a flow-induced density variation
that is not present for Newtonian incompressible fluids. The method outlined
below is designed to take care of this additional density variation.

Consider a uniform spatial grid with lattice spacing $\delta x$. The
LB scheme evolves a set of lattice density distributions $n_i$
by a discrete collision-and-streaming update,
\begin{multline}\label{lbupdate}
  n_i(\vec r+\vec c_i\delta t,t+\delta t)
  =n_i^*(\vec r,t)\\
  =n_i(\vec r,t)+\Delta_i[n(\vec r,t)]+F_i\,,
\end{multline}
where $\delta t$ is the time step of the simulation and the $\vec c_i$ are a
set of discrete velocities, suitably chosen to ensure the desired continuum
limit. The $n_i^*$ are referred to as the post-collision distributions.
$\Delta_i$ is the collision operator (specified later), constructed to
relax the lattice densities to a suitably chosen equilibrium distribution
that restores the Newtonian-fluid case in the continuum limit.
$F_i$ is a driving term and can be constructed to account for the external
body-force density, although this is a nontrivial matter
\cite{Ladd.2001,Nash.2008}.
We will also use it to include non-Newtonian stresses.

To clarify the choice of $F_i$, we briefly
repeat the main points of the Chapman-Enskog expansion describing the
continuum limit of the LB scheme, following the presentation by D\"unweg and
Ladd \cite{Duenweg.2009}. Introducing a small parameter $\varepsilon$ (the
Knudsen number), we set $\tilde{\vec r}=\varepsilon\vec r$, $\tilde
t=\varepsilon t$, and $\hat t=\varepsilon^2t$. We assume an expansion of the
distribution functions $n_i=n_i^{(0)}+\varepsilon n_i^{(1)}+\cdots$, and
equivalently for $\Delta_i$ and $F_i$. The LB update can then be discussed on
the two time scales of momentum convection and diffusion. To zeroth order
in $\varepsilon$, we get $\Delta_i^{(0)}=F_i^{(0)}=0$; the $n_i^{(0)}$ are
collision invariants and thus identified with the equilibrium distribution
$n_i^\text{eq}$. It should be a function of $\rho$ and $\vec\jmath$ only,
in order to avoid spurious invariants of the algorithm.

From the first two orders of
Eq.~\eqref{lbupdate}, one gets
\begin{subequations}\label{lbce}
\begin{gather}\label{lbce1}
  \vec c_i\cdot\vec\partial_{\tilde r}n_i^{(0)}+\partial_{\tilde t}n_i^{(0)}
  =(1/\delta t)\Delta_i^{(1)}+(1/\delta t)F_i^{(1)}\,,
  \\
  \begin{split}
  \frac12\left[(\vec c_i\cdot\vec\partial_{\tilde r})+\partial_{\tilde t}\right]
  \left(n_i^{(1)*}+n_i^{(1)}\right)+\partial_{\hat t}n_i^{(0)}
  \\
  =(1/\delta t)(\Delta_i^{(2)}+F_i^{(2)})\,.
  \end{split}
\end{gather}
\end{subequations}
Recall now that $\vec\partial=\varepsilon\vec\partial_{\tilde r}$ and
$\partial_t=\varepsilon\partial_{\tilde t}+\varepsilon^2\partial_{\hat t}$.
Summing over all directions, i.e., taking the zeroth velocity-moment of
Eqs.~\eqref{lbce}, these equations can be combined to give
$\partial_t\sum_i(n_i^*+n_i)/2+\vec\partial\cdot\sum_i\vec c_i(n_i^*+n_i)/2
=(\varepsilon/\delta t)\sum_i(\Delta_i^{(1)}+F_i^{(1)})+O(\varepsilon^2)$.
The continuity equation for the density,
Eq.~\eqref{continuityrho}, is recovered if the term on the right-hand side
vanishes identically (which is the case in standard LB schemes),
or can be written as a time-derivative. In order to
account for non-Newtonian pressure changes, we adopt the second choice.
Specifically, set $(\varepsilon/\delta t)\sum_i(\Delta_i^{(1)}+F_i^{(1)})
=(\varepsilon/\delta t)(\delta\rho(t)-\delta\rho(t-\delta t))
=\partial_t\delta\rho+O(\varepsilon)$, where
$\delta\rho=-\tr\boldsymbol\Sigma^\text{nN}/(c_s^2d)$ is the
flow-induced density change. We then define the hydrodynamic density
and current as
\begin{subequations}\label{rhoj}
\begin{gather}
  \rho(\vec r,t)=\sum_in_i^\text{eq}
  =\sum_i\frac{n_i^*(\vec r,t)+n_i(\vec r,t)}{2}
  -\delta\rho(\vec r,t)\,,\\
  \vec\jmath(\vec r,t)=\sum_i\vec c_in_i^\text{eq}
  =\sum_i\vec c_i\frac{n_i^*(\vec r,t)
  +n_i(\vec r,t)}{2}\,.
\end{gather}
\end{subequations}
This definition of the density reduces to the standard LB expression
$\sum_in_i$ in the case of traceless non-Newtonian stresses.
As we shall see below, our definition of the momentum current agrees
with the standard one used in LB simulations with external force densities
\cite{Guo.2002}, and it reduces to $\sum_i\vec c_in_i$ in the case
of vanishing external force.

Taking the first velocity-moment of Eqs.~\eqref{lbce} results in
\begin{equation}\label{lbns}
  \partial_tj_\alpha+\partial_\beta\pi_{\alpha\beta}^\text{eq}
  +\frac12\partial_\beta\left(\pi^{\text{neq}*}_{\alpha\beta}
  +\pi^\text{neq}_{\alpha\beta}\right)=f_\alpha^\text{ex}
\end{equation}
where we have introduced the second moment of the distribution
functions, $\pi_{\alpha\beta}=\sum_ic_{i\alpha}c_{i\beta}n_i$
and set $n^\text{neq}=\varepsilon n^{(1)}+O(\varepsilon^2)$.
We have also set
$f_\alpha^\text{ex}=\sum_ic_{i\alpha}(\Delta_i+F_i)/(\delta t)$, which will
be identified with the external body-force density.
Equation~\eqref{lbns} shall recover the Navier-Stokes equation,
Eq.~\eqref{ns}.
Since both the equilibrium pressure $p_0$ and the
Eulerian stresses are given by $\rho$ and $\vec u$ alone, they should
be connected to the equilibrium contribution in Eq.~\eqref{lbns},
\begin{equation}\label{pieq}
  \pi^\text{eq}_{\alpha\beta}=p_0\delta_{\alpha\beta}+\rho u_\alpha u_\beta\,.
\end{equation}

Taking Eqs.~\eqref{rhoj} and \eqref{pieq} together, a second-order
accurate expression for the LB equilibrium distribution is
\begin{equation}
  n_i^\text{eq}(\rho,\vec u)=a^{c_i}\rho\left(1+\frac{\vec u\cdot\vec c_i}
  {c_s^2}+\frac{(c_{i\alpha}c_{i\beta}-c_s^2\delta_{\alpha\beta})
  u_{i\alpha}u_{i\beta}
  }{2c_s^4}\right)
\end{equation}
if the set of $\vec c_i$ obeys the symmetries of a cubic lattice and
at least three different magnitudes $c_i$ are chosen. The weight
coefficients $a^{c_i}$ depend only on these magnitudes and
obey $\sum_ia^{c_i}=1$. For our 2D calculations, the simplest set of
velocities involves nine vectors, termed the D2Q9 model:
$\vec c_0=(0,0)$, $\vec c_{1\ldots 4}/c=(\pm1,0),(0,\pm1)$, and
$\vec c_{5\ldots8}/c=(\pm1,\pm1)$, with weights
$a^0=4/9$, $a^1=1/9$, and $a^{\sqrt2}=1/36$. Here, $c=(\delta x)/(\delta t)$
is the lattice-velocity unit. The speed of sound on the
LB lattice is given by $c_s=c/\sqrt3$, and the equation of state is
the linearized expression for small density variations,
$p_0=\rho c_s^2$, where a constant pressure may be omitted in
incompressible flow.

The nonequilibrium part of the $\pi_{\alpha\beta}$ must recover the
viscous stresses. By comparison of Eqs.~\eqref{ns} and \eqref{lbns},
$\Sigma_{\alpha\beta}
=(\pi^{\text{neq}*}_{\alpha\beta}+\pi^\text{neq}_{\alpha\beta})/2$.
Making use of the equilibrium distribution function, the second moment
of Eq.~\eqref{lbce1},
\begin{equation}\label{lb2}
  \partial_{\tilde t}\pi_{\alpha\beta}^{(0)}+\partial_{\tilde\gamma}
  \phi_{\alpha\beta\gamma}^{(0)}=(1/\delta t)\left(\pi^{(1)*}_{\alpha\beta}
  -\pi^{(1)}_{\alpha\beta}\right)\,,
\end{equation}
can be explicitly calculated. Here, $\phi_{\alpha\beta\gamma}$ is the
third moment of the LB distribution function. From the
conservation laws and the LB equation of state, it follows
$\phi^{(0)}_{\alpha\beta\gamma}=\rho c_s^2(u_\alpha\delta_{\beta\gamma}
+u_\beta\delta_{\alpha\gamma}+u_\gamma\delta_{\alpha\beta})$ and
$\partial_{\tilde t}\pi^{(0)}_{\alpha\beta}=-c_s^2(\delta_{\alpha\beta}
\partial_{\tilde\gamma}j_\gamma+u_\beta\partial_{\tilde\alpha}\rho
+u_\alpha\partial_{\tilde\beta}\rho)+(u_\beta f_\alpha^{\text{ex}(1)}+u_\alpha
f_\beta^{\text{ex}(1)})+(c_s^2\delta_{\alpha\beta}-u_\alpha u_\beta)(\partial_
{\tilde t}\delta\rho)$, where we have neglected terms
of $O(\tlname{Ma}^3)$.
Taking these results together we arrive at
\begin{multline}\label{pineqkappa}
  \pi^{\text{neq}*}_{\alpha\beta}-\pi^{\text{neq}}_{\alpha\beta}
  =(\delta t)\rho c_s^2(\partial_\alpha u_\beta+\partial_\beta u_\alpha)
  \\
  +(\delta t)(u_\alpha f_\beta^\text{ex}+u_\beta f_\alpha^\text{ex})
  \\
  +(c_s^2\delta_{\alpha\beta} -u_\alpha u_\beta)(\delta t)(\partial_t\delta\rho)
  \,.
\end{multline}
The approximations involved in deriving this equation imply that
the LB method provides an approximation of the deviatoric stress tensor
that is second order in the lattice parameters, although this can be
spoiled by the boundary conditions \cite{Krueger.2009,Krueger.2010}.

To proceed further, assume a collision operator of the form
$\Delta_i=\mathcal L_{ij}n_j^\text{neq}$. We specifically employ
the single-relaxation-time BGK model, $\mathcal L_{ij}=-\delta_{ij}/
\tau_\text{LB}$, so that $\pi^{\text{neq}*}_{\alpha\beta}
-\pi^\text{neq}_{\alpha\beta}=-(1/\tau_\text{LB})\pi^\text{neq}_{\alpha\beta}
+\mathcal F_{\alpha\beta}$, where $\mathcal F_{\alpha\beta}=
\sum_ic_{i\alpha}c_{i\beta}F_i$ is the second velocity-moment of the LB
forcing term. Combining this with Eq.~\eqref{pineqkappa} to evaluate
$\pi_{\alpha\beta}^{\text{neq}*}+\pi_{\alpha\beta}^\text{neq}
=(1-2\tau_\text{LB})(\pi^{\text{neq}*}_{\alpha\beta}-\pi^\text{neq}_{\alpha
\beta})+2\tau_\text{LB}\mathcal F_{\alpha\beta}$, we
can by comparison of Eqs.~\eqref{ns} and \eqref{lbns}
identify the viscous stresses: from the first term in Eq.~\eqref{pineqkappa},
we get the Newtonian stresses, and $\mathcal F_{\alpha\beta}$ is still
at our disposal to incorporate non-Newtonian stresses. However,
Eq.~\eqref{pineqkappa} contains spurious terms involving $\vec f^\text{ex}$
and $\delta\rho$. To cancel them, $\mathcal F_{\alpha\beta}$ must contain
matching terms, leaving us with
\begin{multline}\label{lbfab}
  {\mathcal F}_{\alpha\beta}=\left(1-{\textstyle\frac1{2\tau_\text{LB}}}\right)
  (\delta t)(u_\alpha f_\beta^\text{ex}+u_\beta f_\alpha^\text{ex})\\
  +\left(1-{\textstyle\frac1{2\tau_\text{LB}}}\right)
  (\delta t)(c_s^2\delta_{\alpha\beta}-u_\alpha u_\beta)(\partial_t\delta\rho)
  -\frac1{\tau_\text{LB}}\bar\Sigma_{\alpha\beta}^\text{nN}
\end{multline}
The Newtonian viscosity is given as usual for the BGK collision model
\cite{Duenweg.2009},
\begin{equation}\label{taulb}
  \eta^\text{N}=(\delta t)\rho c_s^2(\tau_\text{LB}-1/2)\,.
\end{equation}

Equation \eqref{lbfab}
is a condition on the second moments of the LB forcing term.
From our definitions of $f^\text{ex}_\alpha$ and $\partial_t\delta\rho$,
we get further conditions on the first and zeroth velocity-moments of $F_i$,
viz.\ $\sum_ic_{i\alpha}F_i=(\delta t)f^\text{ex}_\alpha(1-1/(2\tau_\text{LB}))$
and $\sum_iF_i=(\delta t)(\partial_t\delta\rho)(1-1/(2\tau_\text{LB}))$.
This allows to construct $F_i$ with second-order accuracy,
\begin{align}\label{bodyforce}
  F_i&=a^{c_i}\bigg\{\frac{-1}{2c_s^4\tau_\text{LB}}\bar\Sigma_{\alpha\beta}
  ^\text{nN}
  (c_{i\alpha}c_{i\beta}-c_s^2\delta_{\alpha\beta})\nonumber\\
&+\Big(1-\frac 1 {2\tau_\text{LB}}\Big)\Big[(\delta t)(\partial_t\delta\rho)+\frac{f_\alpha^\text{ex}
  c_{i\alpha}}{c_s^2}+\frac{c_{i\alpha}c_{i\beta}-c_s^2\delta_{\alpha\beta}}{2c_s^4}\times\nonumber\\
&\Big(-(\delta t)(\partial_t\delta\rho)u_\alpha
    u_\beta+(u_\beta f_\alpha^\text{ex} + u_\alpha f_\beta^\text{ex}
  )\Big)\Big]\bigg\}
\end{align}
In the case where there is no external body force and no flow-induced
pressure, only the first line in this expression remains, providing a simple
way to incorporate non-Newtonian deviatoric stresses in the LB algorithm.
As a cross-check, imagine a LB simulation with $\tau_\text{LB}$ set up
according to some Newtonian viscosity $\eta$; and set
$\Sigma_{\alpha\beta}^\text{nN}=\Delta\eta(\partial_\alpha u_\beta
+\partial_\beta u_\alpha)$ to be of Newtonian form again.
Reconstructing $n_i^\text{neq}$ up to
second order through its known first moments, direct inspection reveals
$\Delta_i+F_i=-(1/\tau'_\text{LB})n_i^\text{neq}$, where
$\tau'_\text{LB}$ corresponds to Eq.~\eqref{taulb} with
$\eta\mapsto\eta+\Delta\eta$.

Treating $\delta\rho\propto-\delta p^\text{nN}$ explicitly is a way
to incorporate the advection of flow-induced pressure in the scheme.
For the simulations of channel flow driven
by a body force, our choice of normal-stress advection,
\begin{align}
  \label{rhoplus}
\partial_t\delta\rho(t)=-\frac 1 {c_s^2 d\delta t}\left(\Sigma_{\gamma\gamma}^\text{nN}(t)-\Sigma_{\gamma\gamma}^\text{nN}(t-\delta
  t)\right)\,,
\end{align}
allows the fluid to build up a pressure gradient countering the normal stresses
acting on the plug. If we did not take care of the non-Newtonian pressure
contributions in this way, the effect of normal stresses would be lost in a
local change of the bulk viscosity of the fluid.


\subsection{Boundary Conditions}

To maintain a constant pressure gradient along a channel of length $L$,
we employ generalized periodic boundary counditions (GPBC) proposed
by Kim and Pitsch \cite{Kim.2007}. One exploits the simple equation of
state of the LB scheme and identifies the average density of the
outgoing lattice populations at the low-pressure end of the channel with
that of the incoming ones at the high-pressure end, suitably scaled.
Specifically, denote by $p_\pm$ the inlet and outlet pressures, and
by $\bar\rho_\text{in/out}$ the average values of the lattice densities at the
inlet and outlet columns, averaged over the transverse spatial directions.
One then sets
\begin{equation}
  n_i^*(\rho,\vec\jmath)|_\text{in}=n_i^\text{eq}(c_s^{-2}p_++\rho_\text{out}
  -\bar\rho_\text{out},\vec\jmath)+n_i^{\text{neq}*}|_\text{out}
\end{equation}
for each inlet node and direction $\vec c_i$ connecting to the correspoding
outlet node, and vice versa. The construction copies over any density
fluctuations in the transverse directions, only adusting the average
densities.

Keeping the inlet and outlet pressures fixed negates the effect of
implementing the non-Newtonian pressure $\delta\rho\, c_s^2$. Therefore, when using
GPBC, it proves sufficient to construct a body-force term with vanishing
zeroth and first moment, keeping the first line only of Eq.~\eqref{bodyforce}.
Note that because the $F_i$ take the traceless stress tensor,
still a non-trivial pressure gradient in transversal flow direction
emerges. In steady flows, this is easy to see, as Eq.~\eqref{rhoplus} vanishes
and both, the full and the shortened body-force terms, share the same long
time limit. For the cases we consider below, the two methods (GPBC without
density correction, or body force with correction) yield identical results
up to a reinterpretation of the LB density.

The GPBC for pressure gradients can
be extended to sudden pressure jumps. In the case of switch-on,
we pre-initialize the LB densities with a linear gradient to minimize
lattice oscillations due to an unphysical LB shock wave.

The simulated channel is bounded by hard walls in the transverse
direction(s). For their treatment, we employ simple bounce-back
boundary conditions \cite{Succi}. In some cases, we also compare with
velocity-driven planar Couette flow, where the boundary condition at the
moving wall was implemented following Zou and He \cite{Zou.1997}.

\subsection{Numerical Details}

In the following calculations, we adjust $\tau_\text{LB}\approx0.9$,
which is close to the optimum reducing the error in the shear stress
of a Newtonian fluid \cite{Holdych.2004,Krueger.2009}.
We use a grid of $200\times20$
lattice nodes for $\theta\le100$. Higher $\theta$ require to resolve larger
viscosities and viscosity differences, so that we increase the resolution to
$400\times20$ nodes for $\theta=1000$. The scheme was implemented in the
open-source lattice Boltzmann code Palabos \cite{palabos}.

To evaluate the constitutive equation, we will typically need to evaluate
the velocity-gradient tensor $\bs\kappa(t)=(\vec\partial\vec u(t))^T$.
It has in general to be evaluated using a finite-difference scheme on the
LB lattice. In case the constitutive equation can fully be specified
in terms of the symmetric velocity gradients, $\bs D(t)=\bs\kappa(t)
+\bs\kappa^T(t)$, one can make use of Eq.~\eqref{pineqkappa} and evaluate
$\bs D(t)$ directly from the nonequilibrium distributions on a single
lattice node. This method of evaluating the symmetric velocity gradients
is second-order accurate \cite{Boyd.2006}, while a simple finite-difference
scheme is only accurate to first order \cite{Gabbanelli.2005}.
However, since $\pi^\text{neq}_{\alpha\beta}$ contains
$\mathcal S_{\alpha\beta}(\dot\gamma)$ again, the determination of
$\bs D$ from Eq.~\eqref{pineqkappa} turns into an implicit equation that
would generally have to be solved by iteration. Furthermore, we will
consider a constitutive equation that does not depend on $\bs D(t)$ only.
We therefore use
second-order accurate three-point finite differences to evaluate
the shear-rate tensor.

Let us briefly discuss the relation to previous approaches in modeling
non-Newtonian fluids with LB. The most common approach is to adapt
$\tau_\text{LB}(\dot\gamma)$ according to Eq.~\eqref{taulb}, where the
Newtonian viscosity is replaced with a given $\eta_\text{eff}(\dot\gamma)$,
and to set $F_i=0$.
The fluid then remains locally Newtonian. As a consequence, tensorial
aspects such as the normal stress differences are not taken into account
in this class of generalized Newtonian fluids.
Furthermore, the LB method becomes unstable if $\tau_\text{LB}$ drops
to $1/2$, and it works best if the relaxation parameter is chosen within
some bounds close to unity. Constitutive equations that lead
to strong deviations from these bounds are potentially problematic in this
approach.

Other methods, close relatives of lattice kinetic schemes, modify the
LB-equilibrium distribution function $n_i^\text{eq}$. In its second moment,
a non-Newtonian stress contribution is included.
In the steady state, we did not find significant differences to our scheme,
which we believe to be easier to justify in the non-steady case.
There are applications of LB to complex fluids where
the route using a forcing term $F_i$ is empirically found to be more robust 
compared to the modification of the distribution function
\cite{Sulaiman.2006}.

A number of schemes exploit the equivalence of
$\vec\partial\cdot\bs\Sigma$ with an external force density in
Eq.~\eqref{ns}. This seems to be particularly useful if the scheme to
solve the constitutive equation entails evaluation of the stress gradients.
For the present case, we found the approach outlined above to be somewhat
simpler, and we do not expect significant differences.

\section{Nonlinear Maxwell Model}\label{maxwell-model}

\subsection{Constitutive Equations}

To arrive at a constitutive equation for the flow,
we start from a generalized Green-Kubo relation for the nonlinear-response
shear stress
that has been worked out in the ITT formalism. We assume some general
time-dependent incompressible flow, described by the velocity-gradient
tensor $\bs\kappa(t)=(\vec\partial\vec u)^T$; incompressibility implies
$\tr\bs\kappa=0$. We also neglect the advection of stress gradients.
If the flow is switched on at $t=0$ in an equilibrated quiescent system,
\begin{equation}\label{gk}
  \bs\Sigma(t)=\int_0^t\left[-\partial_{t'}\bs B(t,t')\right]
  G(t,t',[\bs\kappa])\,dt'\,,
\end{equation}
where $\bs B(t,t')=\bs E(t,t')\bs E^T(t,t')$ is the Finger tensor,
given by the deformation tensor $\bs E(t,t')$ that describes the transformation
of a material vector $\vec r'$ at an earlier time $t'$ to a vector
$\vec r$ at some later time $t$. The deformation tensor obeys
$\partial_t\bs E(t,t')=\bs\kappa(t)\bs E(t,t')$ and
$\partial_{t'}\bs E(t,t')=-\bs E(t,t')\bs\kappa(t')$, with initial
condition $\bs E(t,t)=\bs1$.

The function $G(t,t',[\bs\kappa])$ is a generalized dynamical shear modulus,
given microscopically as a stress-stress autocorrelation function.
It will in general depend on two time arguments $t$ and $t'$ separately,
while in steady state this dependence reduces to one on $t-t'$ only.
The third argument indicates a dependence on the full flow history at
all previous times $t'\le t$. In linear response, this dependence can be
neglected, but it is essential to describe non-Newtonian fluids.
The principle of material objectivity suggests that the flow history
enters the dynamical shear modulus only through the invariants of the
Finger tensor (as in the schematic model of Ref.~\onlinecite{Brader.2009}),
or through invariants of the symmetrized shear-rate tensor $\boldsymbol D$
(the simplified case considered below).

In quiescent dense liquids, $G(t,t')$ typically decays on a slow structural
relaxation time scale $\tau$ much larger than the microscopic time scale
$\tau_0$. Since $\tau\gg\tau_0$, visco-elastic effects arise in a large
intermediate time window.
We consider flows of some characteristic rate $\dot\gamma$,
where the dressed P\'eclet number $\tlname{Pe}=\dot\gamma\tau\gg1$, but
the bare $\tlname{Pe}_0=\dot\gamma\tau_0\ll1$. It is then convenient to
model the short-time contributions to the viscosity as quasi-instantaneous,
setting
\begin{equation}\label{Sigmamodel}
  \bs\Sigma(t)=\bs\sigma(t)+\eta_\infty\bs D(t)\,.
\end{equation}
Formally, this is achieved by assuming $G(t,t')=G_\text{micr}(t,t')
+G_\text{struc}(t,t')$ to consist of a
slowly relaxing structural part $G_\text{struc}(t,t')$ obeying
$G_\text{struc}(t,t)=G_\infty$, and a short-time
contribution $G_\text{micr}(t,t')\approx(G_0-G_\infty)\,\Theta(\epsilon-(t-t'))$
modeled by a Heaviside function. Inserting into Eq.~\eqref{gk} and
taking the limit $\epsilon\to0$,
we can identify $(G_0-G_\infty)\epsilon=\eta_\infty=:G_\infty\tau_0$
as the short-time Newtonian viscosity.
In rheological terms, $G_0$ corresponds to the high-frequency shear modulus
probed at $(t-t')\ll\tau_0$, while $G_\infty$ is the low-frequency
Maxwell plateau modulus.
Note that $\boldsymbol\sigma$ in Eq.~\eqref{Sigmamodel} is not
necessarily traceless.

From the time derivative of Eq.~\eqref{gk}, one obtains
\begin{multline}\label{gkder}
  \dot{\bs\sigma}(t)-\bs\kappa(t)\cdot\bs\sigma(t)
  -\bs\sigma(t)\cdot\bs\kappa^T(t)
  =\bs D(t)G_\infty\\
  +\int_0^t\left[-\partial_{t'}\bs B(t,t')\right]
  \partial_tG(t,t',[\bs\kappa])\,dt'\,,
\end{multline}
where we have dropped the subscript on $G_\text{struc}$ for convenience.
The terms on the left-hand side are the upper-convected derivative
of the stress tensor.

Assume now steady-state flow for a class of generalized Maxwell models, where
$\partial_tG(t,t',[\bs\kappa])=(-1/\tau_M[\bs\kappa])G(t-t',[\bs\kappa])$
with a relaxation time that is allowed to depend on the deformation rate.
The integral on the r.h.s.\ of the above equation then yields
$-\bs\sigma/\tau_M$, and using $\dot{\bs\sigma}=\bs0$ we arrive at the
formal steady-state solution
\begin{subequations}\label{steadystatemodel}
\begin{equation}\label{sigtsum}
  \bs\sigma_\text{ss}=\sum_{n\ge1}G_\infty\tau_M^n\bs d^n
\end{equation}
where we have defined symmetric matrices
\begin{equation}\label{smalld}
  \bs d^n:=
    \sum_{m=0}^n\begin{pmatrix}n\\ m\end{pmatrix}
    \bs\kappa^m\cdot{\bs\kappa^T}^{n-m}
\end{equation}
\end{subequations}
obeying $\bs d^0=\bs1$, $\bs d^1=\bs D$, and $\bs\kappa\bs d^n
+\bs d^n\bs\kappa^T=\bs d^{n+1}$.
Note that for flows with constant geometry, $\bs d^n=\partial_t^n
\bs B(t,t')|_{t'=t}$.
In general, it is nontrivial to ensure that the infinite sum \eqref{sigtsum}
converges.
For simple shear flow, $\kappa_{\alpha\beta}
=\dot\gamma\delta_{\alpha x}\delta_{\beta y}$, we have
$\bs\kappa\cdot\bs\kappa=\bs0$, so that $\bs d^2=2\bs\kappa\cdot\bs\kappa^T$
and all terms for $n>2$ vanish. The model then contains a shear stress,
$\sigma_{xy}=G_\infty\dot\gamma\tau_M$, and a first normal-stress
difference $N_1=\sigma_{xx}-\sigma_{yy}=2G_\infty\dot\gamma^2\tau_M^2$.
The two quantities obey a simple scaling, $G_\infty N_1/\sigma_{xy}^2=2$.
Since the other diagonal elements of the stress tensor vanish, we also
have $\delta p=-(1/d)\sigma_{xx}=-(2/d)G_\infty\dot\gamma^2\tau_M^2$ (where
$d$ denotes the spatial dimension).

Equation~\eqref{sigtsum} with constant $\tau_M\equiv\tau$ defines the
upper-convected Maxwell model (UCM) \cite{Oswald}.
In simple shear, for $\dot\gamma\tau\ll1$
one obtains a low-shear Newtonian viscosity $\eta=G_\infty\tau$.
For some choices of $\bs\kappa$, Eq.~\eqref{sigtsum} diverges in the UCM:
in planar extensional flow, we have $\kappa_{xx}=-\kappa_{yy}=\dot\epsilon$
and all other $\kappa_{\alpha\beta}=0$. Equation~\eqref{sigtsum} then yields
$\sigma_{xx}=2G_\infty\dot\epsilon\tau/(1-2\dot\epsilon\tau)$, valid only for
$\dot\epsilon\tau<1/2$. This is a well-known deficiency of the UCM which
will be cured in the nonlinear model we discuss below.

In shear-thinning fluids, the structural relaxation time $\tau$ interferes
with the time scale set by the external perturbation, $1/\dot\gamma$.
Correlation functions such as $G(t)$ under strong shear hence decay
on a time scale that is the equlibrium relaxation time $\tau$ as long
as shear is weak ($\tlname{Pe}\ll1$),
but shear induced if $\tlname{Pe}\gg1$. In an ad-hoc way, this can
be modeled by letting
\begin{equation}\label{nlmaxwell}
  \tau_M^{-1}=\tau^{-1}+\tlname{II}_D^{1/2}/\gamma_c\,.
\end{equation}
Here, the second invariant of the symmetric velocity-gradient tensor
appears, $\tlname{II}_D=(1/2)\tr\bs D^2$ in incompressible flow.
It represents the square of an instantaneous shear rate: in simple shear,
$\tlname{II}_D=\dot\gamma^2$, while in planar extensional flow,
$\tlname{II}_D=(2\dot\epsilon)^2$.
The coefficient $\gamma_c$ models a typical strain amplitude relevant for
the shear-induced breaking of nearest-neighbor cages. We typically set
$\gamma_c=0.1$ in numerical calculations and require $\gamma_c\ll1$.

The nonlinear generalized Maxwell model defined by Eq.~\eqref{nlmaxwell}
captures the qualitative effects of shear-induced acceleration of
structural dynamics close to the glass transition. Its steady-state
simple-shear properties have been discussed earlier
\cite{FuchsCatesFaraday,FuchsBallauffCSA,Voigtmann.2011}.
In the tensorial generalization
presented here, it remedies the deficiency of the original, linear-response
UCM: choosing $\gamma_c$ small enough, the infinite sum Eq.~\eqref{sigtsum}
can be forced to converge for any given flow-rate tensor.

The nonlinear generalized Maxwell model incorporates a dynamic yield
stress if $\tau\to\infty$:
let us introduce $\bs\kappa(t)=K(t)\bs\kappa_0$ where the flow rate
$K(t)$ can be chosen positive without loss of generality. We assume here
that the geometry of the flow does not change over time.
Then, $\tlname{II}_D=K(t)^2\tlname{II}_{D_0}$ and the constitutive equation
Eq.~\eqref{sigtsum} for $\tau\to\infty$ gives
$\bs\Sigma=\bs\sigma_y+\eta_\infty\bs D$ with
\begin{equation}\label{yieldstress}
  \bs\sigma_y=\sum_{n\ge1}G_\infty\left(\frac{\gamma_c}{\tlname{II}_{D_0}^{1/2}}
  \right)^n\sum_{m=0}^n\begin{pmatrix}n\\ m\end{pmatrix}
  \bs\kappa_0^m{\bs\kappa_0^T}^{n-m}\,.
\end{equation}
Our constitutive equation can hence be classified as that of a
Bingham fluid in the limit $\tau\to\infty$.


Beyond the steady state, one cannot expect Eq.~\eqref{gkder} to reduce
to a differential equation. While one could consider
$\partial_tG(t,t')=(-1/\tau_M(t))G(t,t')$ together with Eq.~\eqref{nlmaxwell}
as a time-dependent generalization of the model, this would imply that the
dynamical shear modulus depends on the flow only through an accumulated strain
$\int_{t'}^t\tlname{II}_D^{1/2}(s)\,ds$. In cases like large-amplitude
oscillatory shear (LAOS), this does not appear to be plausible, since it
implies that memory effects can ``come back'' after a cycle of strong
shear. It can be argued that in
dense liquids under strong flow, the instantaneous rate
$\tlname{II}_D^{1/2}$ itself should
control a decay rate. In a class of constitutive equations known
as fluidity models \cite{Derec.2001,Coussot.2002,Fielding.2012}, one
essentially imposes $\tau_M(t)$ to obey another differential equation.
Another possibility is to consider a flow dependence of $\tau_M$ that
is not instantaneous in the time $t$, but influenced by earlier times $t'$
\cite{Siebenbuerger.2012,Voigtmann.2013pre}. This results in an
integral generalized-Maxwell models that no longer allows the reduction of
Eq.~\eqref{gk} to a differential constitutive equation.

Such integral constitutive equations can be treated with the algorithm
outlined above \cite{Papenkort.2013phd}, at the cost of much higher
computational power. In the present paper, we will restrict ourselves to the
generalized Maxwell model defined by Eqs.~\eqref{steadystatemodel}
and \eqref{nlmaxwell}, where $\boldsymbol\kappa$ and hence $\tau_M$ are
replaced by their time-dependent, instantaneous values. This instantaneous
nonlinear Maxwell (inlM) model amounts to
focussing on the physics of a yield-stress fluid, and neglects effects
of visco-elasticity.

\subsection{Channel Flow}

The stationary laminar velocity profile in a pressure driven 2D channel flow
of a non-Newtonian fluid described by the nonlinear Maxwell model introduced
above can be calculated analytically. This solution will serve as a useful
reference case to check the LB scheme. Consider a channel of width $2H$
(taken in the $y$-direction) and a pressure drop $\Delta p$ per unit length
($x$-direction).  We introduce
dimensionless quantities by $s=t/\tau_0$ and $d=y/H$; the streaming velocity
of interest then is $v=u_x\tau_0/h$.
We assume spatial homogeneity along the flow direction. In incompressible
flow, $u_y$ then has to vanish identically, and the Navier-Stokes equation
combined with the inlM model reduces to
\begin{equation}\label{nspoiseuille}
  \tlname{Re}\partial_sv=\wp\,H+\partial_d\left[\left(1+\frac{\theta}
  {1+\theta|\partial_dv|/\gamma_c}\right)\partial_dv\right]
\end{equation}
with $\wp=\Delta p/G_\infty$.
$\tlname{Re}=\rho H^2/(\eta_\infty\tau_0)$ is the (worst-case)
Reynolds number of the problem.
The parameter $\theta=\tau/\tau_0$ quantifies the relative enhancement of
the low-shear viscosity over the Newtonian high-shear one; $\theta=0$
corresponds to a Newtonian fluid. As the glass transition is approached,
$\theta\to\infty$.

We further assume no-slip boundary conditions imposed at $d=\pm1$.
Symmetry dictates $\partial_dv=0$ for $d=0$, and we anticipate that
the velocity gradient does not change sign in either half of the channel.
Under these conditions, the flow in the upper half of the channel, in steady
state, is given by a quadratic equation for $\partial_dv$,
\begin{equation}
  0=\wp Hd+(1+\theta-\wp Hd\,\theta/\gamma_c)\partial_dv-(\theta/\gamma_c)
  (\partial_dv)^2
\end{equation}
which is solved by
\begin{multline}\label{maxwellpoiseuille}
  \partial_dv(d)=-\frac{\wp Hd}2+\frac{\gamma_c(1+\theta)}{2\theta}
  \\
  -\sqrt{\left(\frac{\wp Hd}2-\frac{\gamma_c(1+\theta)}{2\theta}\right)^2
  +\frac{\wp Hd\,\gamma_c}{\theta}}\,.
\end{multline}
To obtain the dimensionless velocity $v(d)$, this expression is to be
intregrated on $d\in[0,1]$ and shifted so that $v(1)=0$ is obeyed.

For small $\theta$, the above expression reduces to $\partial_dv(d)
=-\wp Hd(1-\theta)+O(\theta^2)$, so that the Newtonian parabolic velocity profile
expected in Poiseuille flow is recovered. Newtonian profiles are in general
expected (for finite $\theta$) if the pressure is small,
$\partial_dv(d)=-\wp Hd/(1+\theta)+O(\wp^2)$, or large enough to probe the
Newtonian high-shear viscosity, $\partial_dv(d)=-\wp Hd+\gamma_c+O(1/\wp)
\sim-\wp Hd$ as $\wp\gg\gamma_c/(Hd)$.

For $\theta\to\infty$, a singular boundary layer emerges in
Eq.~\eqref{nspoiseuille}, so that one has to distinguish an inner
(center-channel) and outer solution in Eq.~\eqref{maxwellpoiseuille}.
For $d>\gamma_c/(\wp H)$, we get $\partial_dv=-\wp Hd+\gamma_c
+\mathcal O(1/\theta)$, a shifted
parabolic profile. For $d<\gamma_c/(\wp H)$, a ``plug'' solution
$\partial_dv=\mathcal O(1/\theta)$ appears. Specifically,
\begin{equation}\label{plug}
  \partial_dv(d)\sim\begin{cases}
    -\wp Hd+\gamma_c & d>d_c=\gamma_c/(\wp H)\\
    \wp Hd\gamma_c/(pHd-\gamma_c)/\theta & d<d_c\end{cases}\,.
\end{equation}
The expansion in $1/\theta$ is nonanalytic for $d=d_c$, where
the two solutions merge. In particular, we do not find a solution for
$\theta=\infty$, since the model then predicts $\dot\gamma=0$ and
a constant shear stress inside the plug, in violation of the
Navier-Stokes equation $\sigma_{xy}=-\Delta p\, H d$.
The plug boundary is, in line with physical expectation, simply given
by the point where this stress matches the yield stress under shear,
$\sigma_{xy}(d_c)=G_\infty\gamma_c=(\sigma_y)_{xy}$.

According to Eq.~\eqref{sigtsum}, the above steady-state solution implies
a normal-stress difference $N_1=\sigma_{xx}-\sigma_{yy}
=2G_\infty(\partial_yu_x)^2\tau_M^2$,
and a nonequilibrium stress contribution $\delta p=N_1/2$. The
solution with $u_y\equiv0$ is indeed consistent with the Navier-Stokes
equations. In the limit $\theta\to\infty$, there results for the normal
stress difference
\begin{equation}\label{mwnd}
  \sigma_{xx}-\sigma_{yy}\sim\begin{cases}
   2G_\infty\gamma_c^2 & d>d_c\\
   2G_\infty(\wp Hd)^2 & d<d_c\end{cases}\,.
\end{equation}
If the pressure gradient drops
below a certain yield value, $\Delta p<\Delta p_c=2G_\infty\gamma_c/(2H)$,
no flow is found across the entire width of the channel.

The emergence of a yield stress has an interesting consequence for the
temporal evolution of the flow profile after the driving pressure is removed
(or switched to one below the yield value). While a Newtonian fluid's
velocity will decay exponentially in time, the one of a yield-stress
fluid will drop to zero at a finite stopping time
\cite{Huilgol.2002,Huilgol.2002b}.

%

\section{Results}\label{results}

Our constitutive equation becomes Newtonian for both low and high shear
rates: there holds $\boldsymbol\Sigma\sim\eta_\infty\boldsymbol D$ for
$\|\boldsymbol D\|$ large enough, since in this limit $\boldsymbol\sigma$
approaches $\boldsymbol\sigma_y$ which remains bounded. For small
$\|\boldsymbol D\|$ on the other hand, we can set $\tau_M\sim\tau$, and
obtain $\boldsymbol\Sigma=(\eta+\eta_\infty)\boldsymbol D$.

Applying our LB scheme, we have to fix $\tau_\text{LB}$, and consequently
all lattice units, to match a given Newtonian viscosity. Based on the
physical picture, it is tempting to choose the high-shear viscosity
$\eta_\infty$, and hence to identify $\boldsymbol\Sigma^\text{nN}\equiv
\boldsymbol\sigma$. However, we found the accuracy of the LB simulation
to be greatly enhanced by choosing $\tau_\text{LB}$
to match the larger low-shear viscosity of the Maxwell model, identifying
$\boldsymbol\Sigma^\text{nN}=G_\infty(\tau_M-\tau)\boldsymbol D
+\sum_{n\ge2}2G_\infty\tau_M^n\boldsymbol d^n$.
The fact that this contribution becomes negative does not affect the
stability of the scheme as long as the physical viscosity is guaranteed to
be positive.

Here and in the following we choose a pressure difference
$\Delta p=G_\infty/2H$, i.e., a pressure drop comparable to the elastic
modulus of the fluid. This choice is appropriate for soft matter flow and
brings out most clearly the non-Newtonian plug-flow effects.

\subsection{Stationary Profiles}

\begin{figure}
\includegraphics[width=.9\linewidth]{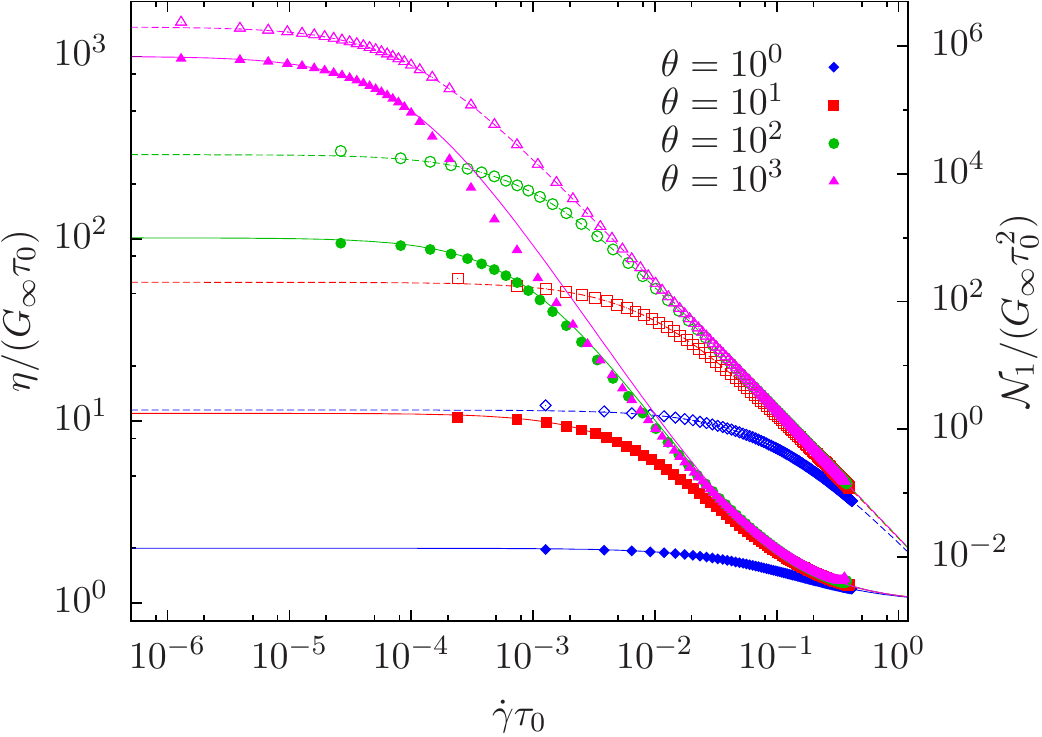}
\caption{\label{flowcurve}
  Shear viscosity $\eta=\sigma_{xy}/\dot\gamma$ (filled symbols; left axis)
  and normal
  stress coefficient $\mathcal N_1=(\sigma_{xx}-\sigma_{yy})/\dot\gamma^2$
  (open symbols;
  right axis) as a function of local shear rate $\dot\gamma$.
  Lines are analytical results from the nonlinear Maxwell model for different
  quiescent relaxation times $\tau=\theta\tau_0$, in units of the
  microscopic relaxation time $\tau_0$;
  symbols are LB simulation results.
}
\end{figure}

The steady state properties of the constitutive equation are illustrated
in Fig.~\ref{flowcurve}. We show the rate-dependent shear viscosity
$\eta(\dot\gamma)=\sigma_{xy}/\dot\gamma$, and the normal stress coefficient
$\mathcal N_1=(\sigma_{xx}-\sigma_{yy})/\dot\gamma^2$, as a function of the
local shear rate $\dot\gamma=\kappa_{xy}$ in the fully developped channel flow.
We will see below that indeed
the steady-state solution is to a very good approximation given by
$u_x(y)$ as the only non-vanishing velocity component, and hence
$\kappa_{xy}$ as the only non-zero velocity gradient.
Since our constitutive equation is local and does not involve gradient
terms, the results shown in Fig.~\ref{flowcurve} are identical to those
obtained in a simple planar shear setup, where a homogeneous shear rate
$\dot\gamma$ is controlled. This has been checked separately.

The nonlinear Maxwell model describes an increase in both shear viscosity
and normal stress coefficient due to slow structural relaxation,
parametrized by the large relaxation time $\tau$. The shear viscosity
for $\dot\gamma\to0$ becomes that of a Newtonian fluid (independent on
shear rate), and grows as a function of $\tau$. Upon increasing shear rate,
as $\dot\gamma\tau\approx\gamma_c$, shear thinning sets in, because
$\tau_M$ is no longer controlled by $\tau$, but by $1/\dot\gamma$.
As a result, $\eta(\dot\gamma)\sim1/\dot\gamma$, i.e., the model contains
the (trivial) shear-thinning exponent $-1$. At large $\dot\gamma$,
the Maxwell contribution to $\eta$ becomes negligible, and a Newtonian
fluid obeying $\eta=\eta_\infty$ results. This regime is not fully
resolved in our channel flow simulations, but is easily accessible to the
LB scheme in simple shear.

At low shear rates, the fluid is not truly Newtonian, since large
normal stress coefficients arise even in incompressible flow.
These scale as $\tau^2$ in the limit
$\dot\gamma\to0$. In the shear-thinning regime, the normal stress
coefficient obeys $N_1\sim1/\dot\gamma^2$ as expected by symmetry --
recall that upon reversing the flow direction, the diagonal elements
of $\boldsymbol\sigma$ do not change sign.

As shown in Fig.~\ref{flowcurve}, our LB scheme (results shown as symbols)
is able to trace the analytical solution of the nonlinear Maxwell model
(lines) over at least six orders of magnitude in the shear
rate, and three orders of magnitude in viscosity variation.
(Implying more than six orders of magnitude change in the normal-stress
coefficient.)
Only at the largest value of $\tau$ considered, $\theta=10^3$, some deviations
can be seen. We expect that a better lattice resolution will improve these
results.

The qualitative features of the flowcurves shown in Fig.~\ref{flowcurve}
are in agreement with many shear-thinning fluids close to a glass transition
for small bare P\'eclet numbers, $\tlname{Pe}_0=\dot\gamma\tau_0\ll1$.
They are also in qualitative agreement with calculations based on a
schematic MCT model \cite{Brader.2009}.
At high $\tlname{Pe}_0$, one usually finds still non-vanishing
normal-stress differences, together with an increasing non-equilibrium
pressure contribution $\delta p$ \cite{Mandal.2012}. This is not captured
in our model, since we assume a purely Newtonian high-shear viscosity.
Augmenting the model to display another Maxwell-type relaxation on the
time scale $\tau_0$ would be closer to experimental and MD simulation
results. In the following, we focus on small shear rates, so that this
difference is not relevant here.

\begin{figure}
\includegraphics[width=.9\linewidth]{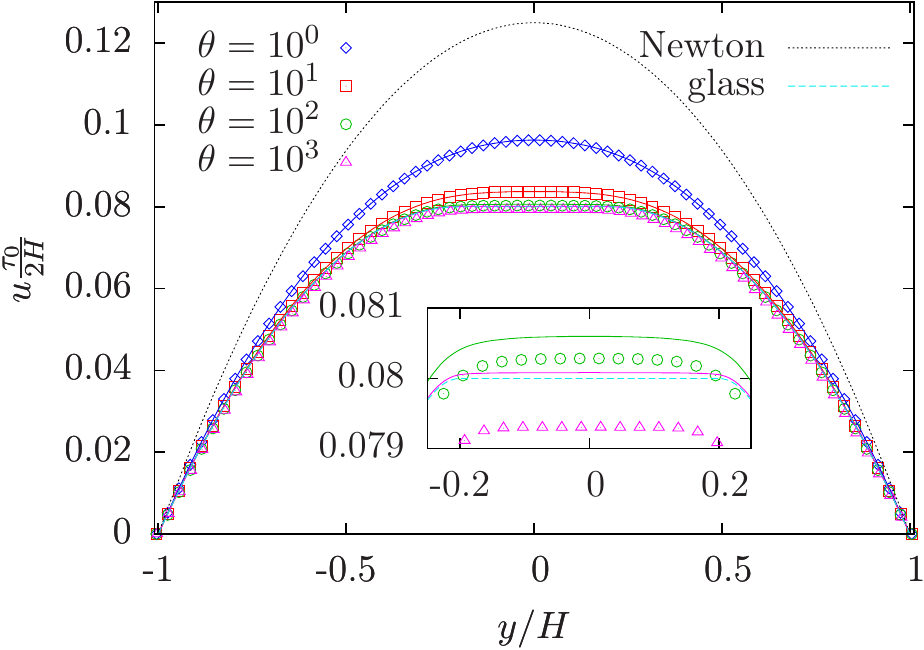}
\caption{\label{profile}
  Steady-state velocity profile in a 2D channel flow driven by a constant
  pressure gradient $\Delta p/(2H)=G_\infty$, for the nonlinear generalized
  Maxwell model, for different $\theta=\tau/\tau_0$.
  Symbols are LB results,
  solid lines are the analytical solutions obtained from
  Eq.~\eqref{maxwellpoiseuille}. A dashed line (see inset) shows the solution for
  $\theta\to\infty$, the dotted line is the parabolic profile for a Newtonian
  fluid, $\theta=0$.
}
\end{figure}

We now turn to the velocity profiles of the planar channel flow.
Figure~\ref{profile} compares the velocity profiles obtained by our LB
simulation to the analytic solution for the strictly incompressible case,
Eq.~\eqref{maxwellpoiseuille}, for various $\theta$. For $\theta\to0$, the
familiar parabolic Pouisseuille flow profile of a Newtonian fluid is
recovered (LB results not shown).
As $\theta$ increases, the center velocities decrease, while the velocity
gradients flatten. As $\theta\to\infty$, a ``plug'' of unsheared liquid
develops in the center of the channel. For $\theta\le10^3$, the LB
results (symbols in Fig.~\ref{profile}) are in good agreement
with the analytic prediction. The largest deviations are seen in the
plug for large $\theta$, as shown in the inset of the figure.
Even for $\theta=10^3$, the relative deviation in the velocity profile
is less than $1\%$. This remaining error is largely governed by the
LB grid resolution.
This is similar to LB simulations where a ``scalar'' constitutive equation
is employed, either in terms of an extra forcing term $F_i$, or through
a local adaption of $\tau_\text{LB}$ as discussed in the introduction.
Using both these schemes for comparison, we found similar errors as the
ones shown in Fig.~\ref{profile}.

The half-width $w$ of the plug follows from Eq.~\eqref{plug},
$w=2G_\infty\gamma_c/(\Delta p\,H)=0.2$, and already
$\theta=10^2$ is very close to the theoretical $\theta\to\infty$ solution
(shown in Fig.~\ref{profile} as a dotted line).
One can view finite $1/\theta$ as a regularization parameter as it
is often employed in numerical calculations involving Bingham or other
yield-stress fluids \cite{Chatzimina.2005}.
In fact, the case $\theta\to\infty$ is an idealization that is not
achieved in reality, as even in the glass, some residual relaxation processes
persist.

\begin{figure}
\includegraphics[width=.9\linewidth]{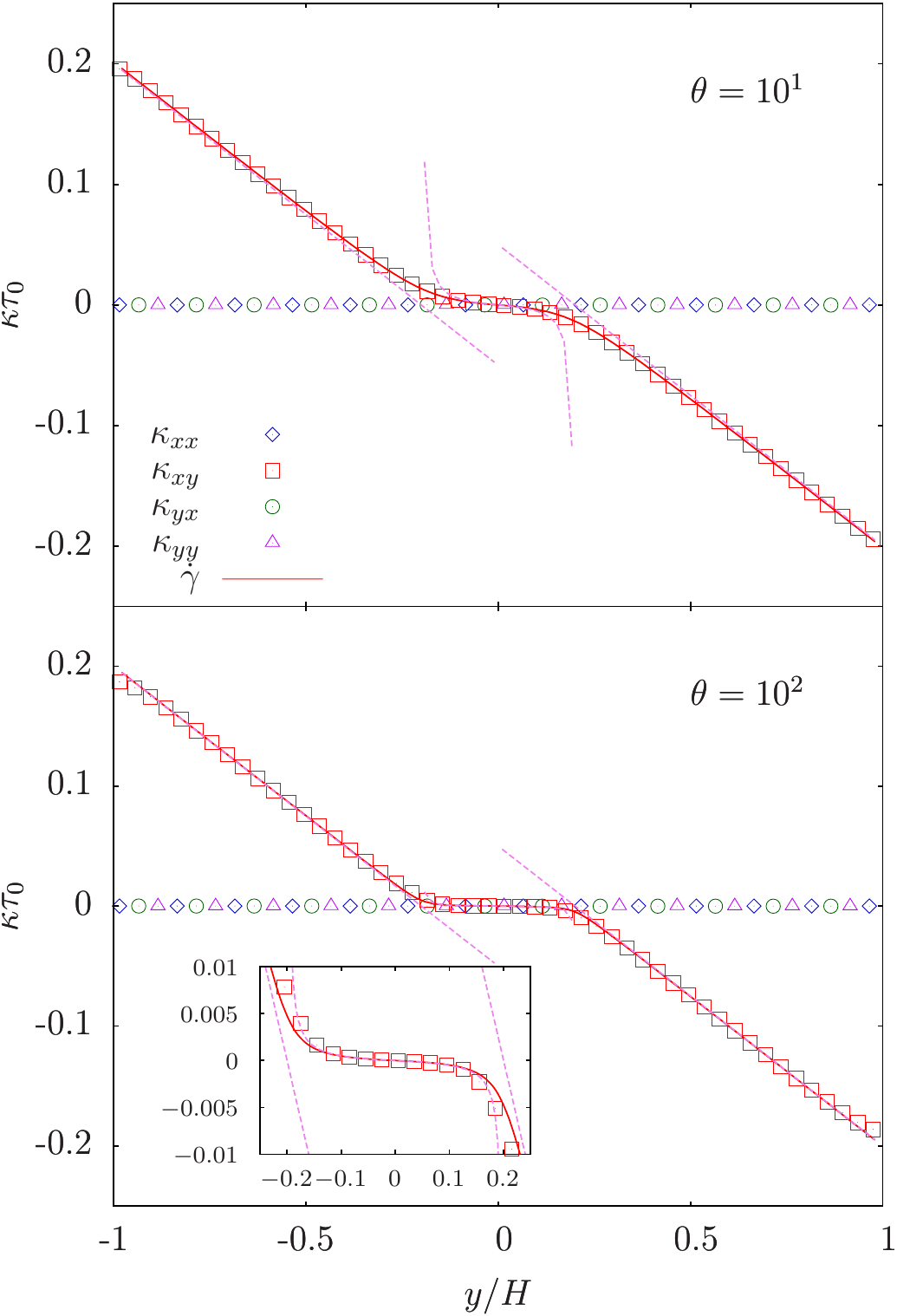}
\caption{\label{channelkappa} Velocity gradient tensor $\boldsymbol\kappa$
  from LB simulations of
  the tensorial generalized Maxwell model, in pressure-driven 2D channel flow.
  The dashed lines show the prediction of the asymptotes Eq.~\eqref{plug}, the
  solid lines the analytic result Eq.~\eqref{maxwellpoiseuille}.  }
\end{figure}

The emergence of a plug region is even more clearly seen in the
velocity gradients. Figure~\ref{channelkappa} shows the elements of
the velocity-gradient tensor $\boldsymbol\kappa$ for the pressure-driven
channel flow discussed in connection with Fig.~\ref{profile}, for
$\theta=10$ and $\theta=100$. The only element that is numerically different
from zero is $\kappa_{xy}$, as expected from the incompressibility
condition.
With our choice of grid parameters, the largest error occurs near the
channel inlet/outlet boundaries, where
$\kappa_{xx}=\mathcal O(5\times10^{-8})$.
Symbols in Fig.~\ref{channelkappa}
show LB simulation results; they agree very well with the
analytical prediction, Eq.~\eqref{maxwellpoiseuille}, shown as solid lines.
Furthermore, even for the moderate $\theta$, the asymptotic result
for $\theta\to\infty$, Eq.~\eqref{plug}, already describes the velocity
gradients surprisingly well (dashed lines).

\begin{figure}
\includegraphics[width=.9\linewidth]{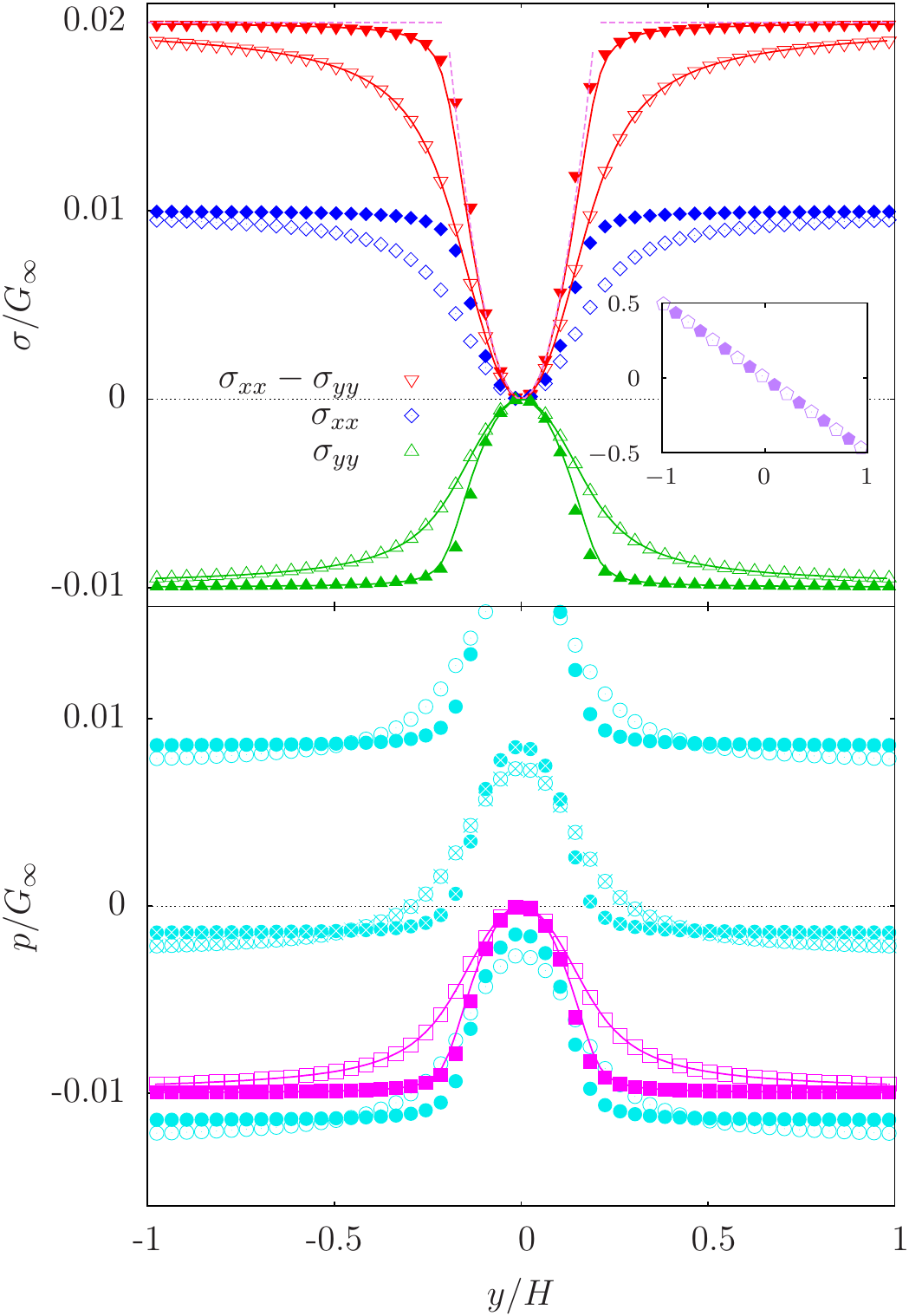}
\caption{\label{channelsigma} Stress tensor elements $\sigma_{\alpha\beta}$
  for pressure-driven channel flow with $\theta=10$ (open symbols) and
  $\theta=100$ (filled symbols). Upper panel:
  Diamonds (triangles) show $\sigma_{xx}$
  ($\sigma_{yy}$), while inverted triangles are the normal stress difference
  from the LB simulation. Lines are the analytical calculation from the
  Maxwell model, assuming incompressible flow.
  The inset shows the expected
  linear behavior of $\sigma_{xy}$.
  Lower panek:
  Circles show different cuts along the channel of the
  flow-induced pressure when using generalized periodic boundary conditions
  including a pressure step along the channel. The center-channel position is
  marked with crosses. If flow is instead driven by a body force, the pressure
  becomes translation-invariant along the channel (squares).
  }
\end{figure}

Figure~\ref{channelsigma} shows the elements of the stress tensor
corresponding to Fig.~\ref{profile} with $\theta=10$ and $\theta=100$.
As shown in the inset, the shear stress $\sigma_{xy}$ obeys the expected
linear behavior dictated by the Navier-Stokes equation. This is the only element
of the stress tensor that is nonzero for the ``scalar'' constitutive
equation incorporated in non-Newtonian LB schemes that adjust
$\tau_\text{LB}$ through iteration.

The normal-stress difference $N_1=\sigma_{xx}-\sigma_{yy}=2G_\infty\dot\gamma^2
\tau_M(\dot\gamma)^2$
contained in the nonlinear Maxwell model can be evaluated easily from
Eq.~\eqref{maxwellpoiseuille}. As demonstrated in
Fig.~\ref{channelsigma}, the LB simulation results (circles) are in excellent
agreement with this prediction (shown as a solid line)
for the values of $\theta$ we investigated.
From Eq.~\eqref{mwnd}, we obtain for
$\theta\to\infty$ a constant normal stress difference outside the plug,
and a parabolic dependence inside. This is shown in the figure as a dashed
line. Already for $\theta=100$, the normal stress coefficient closely
follows this asymptotic prediction.
In Fig.~\ref{channelsigma}, we also show the individual elements of the
deviatoric stress tensor, $\bar\sigma_{xx}$ and $\bar\sigma_{yy}$, obtained
by the LB algorithm. They reconfirm the analytical calculation and highlight
the fact that the modified LB algorithm absorbs the isotropic part of the
non-Newtonian stresses as an additional pressure.

To elucidate this point, we show in the bottom panel of Fig.~\ref{channelsigma}
the overall pressure as a functon of the cross-channel position $y$, for
various cuts at constant $x$-position along the channel. The inclusion of
pressure effects in the nearly-incompressible LB solution is not without
subtlety, and we show two possible approaches: cyan symbols in
Fig.~\ref{channelsigma} correspond to simulations with
generalized periodic boundary conditions incorporating a fixed pressure
difference. Magenta symbols are results obtained with a body force driving
the fluid flow; in this case, the overall pressure is translational-invariant
along the channel. For all other quantities discussed here, the two methods
give results that are numerically indistinguishable; however, for the
case of a body force driving the flow, additional care has to be taken
to account for the non-Newtonian pressure within LB. The generalized
boundary conditions directly control the average pressure and are hence
easier to implement in this case \cite{Papenkort.2013phd}.

The appearance of a positive $N_1$ causes the fluid to be driven towards
the plug, since in the sheared region, forces act perpendicular
to the flow direction towards the center and towards the confining walls.
The latter forces are balanced by the no-flux boundary conditions.
As a result, the pressure inside the plug region increases relative to
the one outside.

\subsection{Transient Dynamics}

We next consider the transient dynamics when going over from the quiescent
state to a flowing steady state, and vice versa, by applying or removing
the pressure difference instantaneously.

\begin{figure}
\includegraphics[width=.9\linewidth]{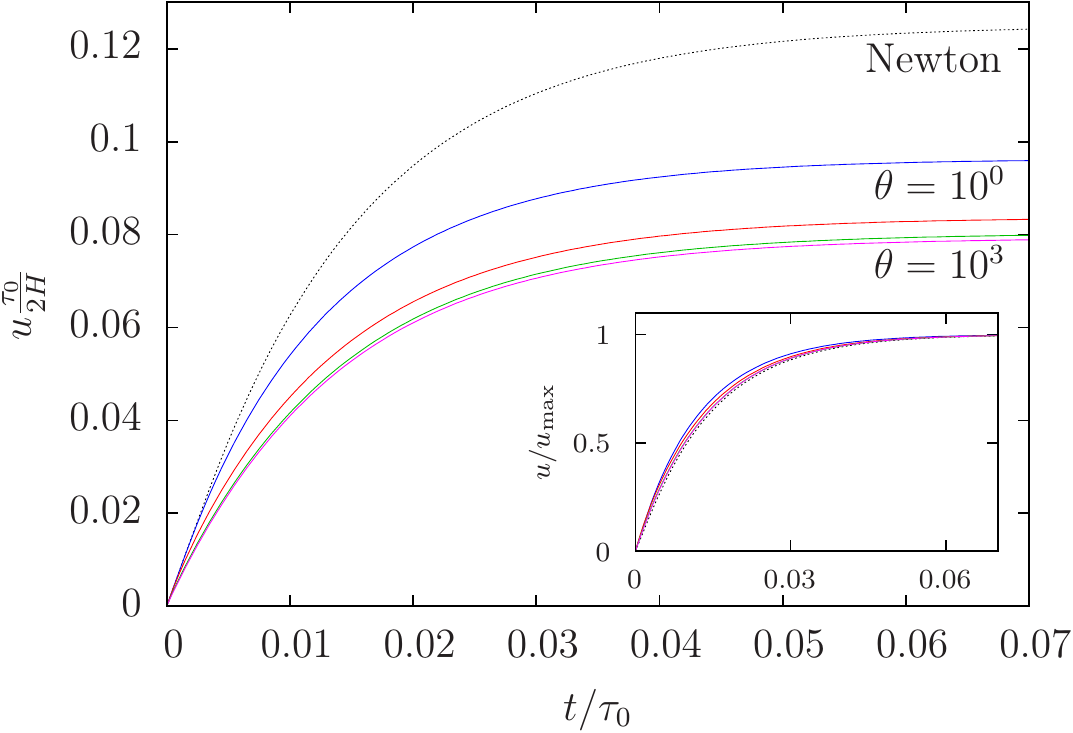}
\caption{\label{start} Evolution of the mid-channel velocity after application
  of a pressure gradient (startup), for different $\theta$. Tensorial Maxwell
  model, LB results. In the inset, the velocities are scaled by the steady
  state value.}
\end{figure}

Figure~\ref{start} shows the evolution of the mid-channel velocity profiles
after startup of 2D channel flow.
In the Newtonian case, an explicit analytical solution is available
\cite{Batchelor}; it essentially shows an exponential increase towards
the steady-state value.
Even for the non-Newtonian case $\theta>0$, no qualitative change is seen.
The switch-on solutions are dominated by equating the time-derivative
of the velocity with the constant pressure-drop term in
Eq.~\eqref{nspoiseuille}, so that the
nonlinear contributions from the stress tensor are small.
As shown in the inset of Fig.~\ref{start}, the startup results can almost
be scaled on top of each other simply by dividing through the steady-state
value.

\begin{figure}
\includegraphics[width=.9\linewidth]{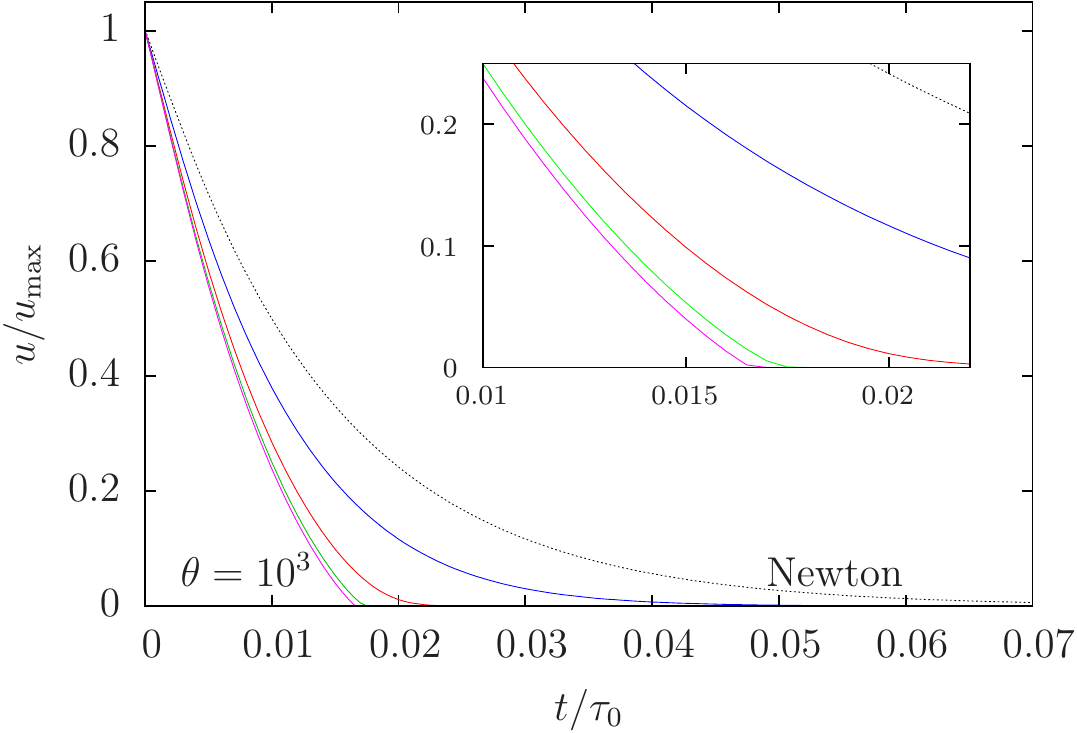}
\caption{\label{stop}
  Evolution of the mid-channel velocity after removal of a constant
  pressure gradient (cessation). Tensorial Maxwell model, LB results.
}
\end{figure}

Figure~\ref{stop} shows the cessation of pressure-driven channel flow
when the applied pressure difference is suddenly removed. Here,
the terms that balance in the Navier-Stokes equation are the time derivative
and the stress-tensor derivative, so that nonlinear contributions to the
latter are much more prominent.
For the Newtonian case, start-up and cessation evolution are symmetric
in the sense that the corresponding results in Figs.~\ref{start} and \ref{stop}
can be collapsed by a simple linear transformation \cite{Batchelor}.
While for any finite $\theta$, the ultimate flow decay is again Newtonian,
for large $\theta$ the cessation profiles indicate the stopping-time
phenomenon quoted above. As evident from Fig.~\ref{stop}, and consistent
with Fig.~\ref{profile}, already $\theta=100$ is representative of the
yield-stress fluid case $\theta=\infty$ in this respect.

The LB algorithm is accurate enough to resolve the finite-time singularity
within reasonable bounds. We estimate a stopping time of
$t\approx0.016\tau_0$. This appears to be in good agreement with the upper bound
estimated by Huilgol \cite{Huilgol.2002}. Previously, the accuracy of this
upper bound was checked in FEM simulations
\cite{Chatzimina.2005,Muravleva.2010}. Our
results indicate that the LB algorithm provides similar accuracy for
non-Newtonian flows.

Note that the appearance of a finite stopping time is a consequence of
the instantaneous nonlinear Maxwell model, or other instantaneous yield-stress
constitutive equations, since it is derived from a variational inequality
that is local in space and time. Incorporation of viscoelastic effects
in the full nonlinear generalized Maxwell model will render the cessation
flow phenomenology more complex.

\begin{figure}
\includegraphics[width=\linewidth]{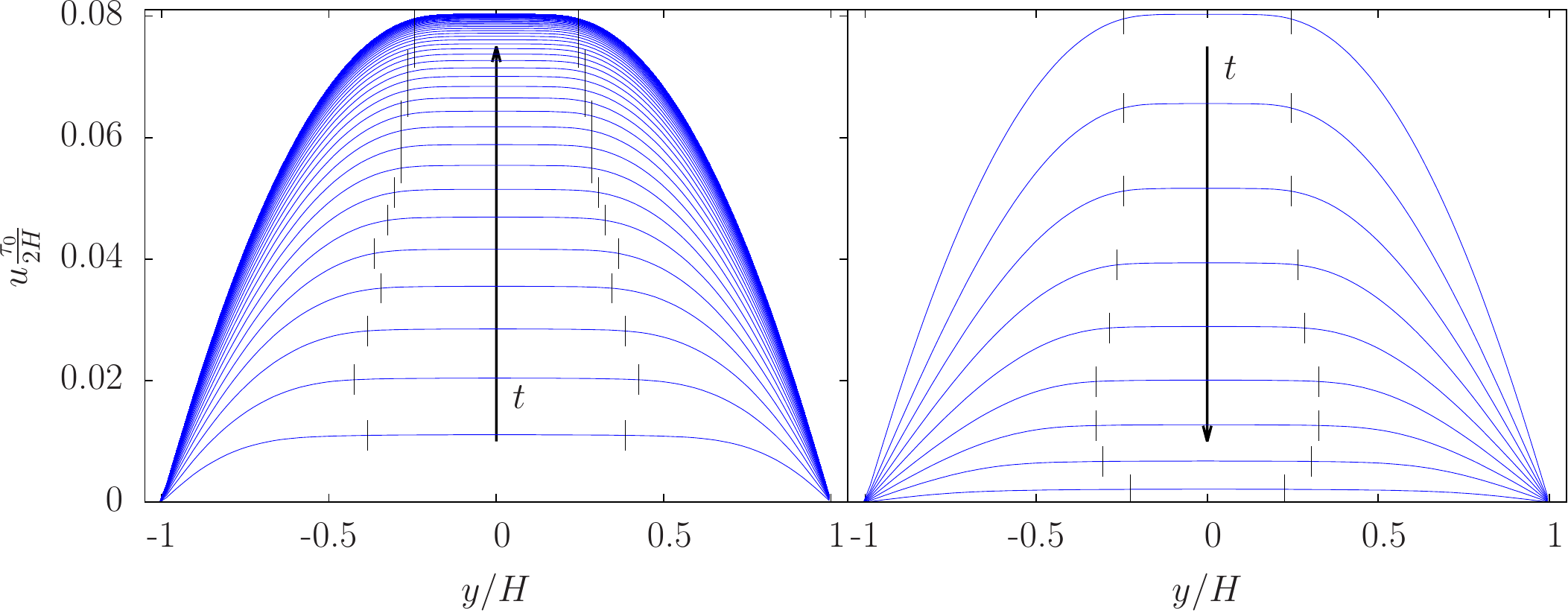}
\caption{\label{stopprofiles}
  Temporal evolution of the velocity after startup (left) respectively
  cessation (right) of pressure-driven 2D channel flow,
  inlM model with $\theta=10^2$.
  Small horizontal lines mark the points where the velocity deviates
  by $1\%$ from the maximum velocity, as an indicator of the plug width.
}
\end{figure}

To complete the picture,
Figure~\ref{stopprofiles} shows the temporal evolution of the plug-flow
profile after application and after removal of the driving pressure gradient.
In startup flow (left panel), the plug-flow profile develops from an
initially flat velocity profile, with a central plug whose width gradually
decreases until it reaches the steady-state width discussed above.
The decay of the velocity profiles after removal of the pressure difference
qualitatively follows the inverse sequence of steps.

\begin{figure}
\includegraphics[width=\linewidth]{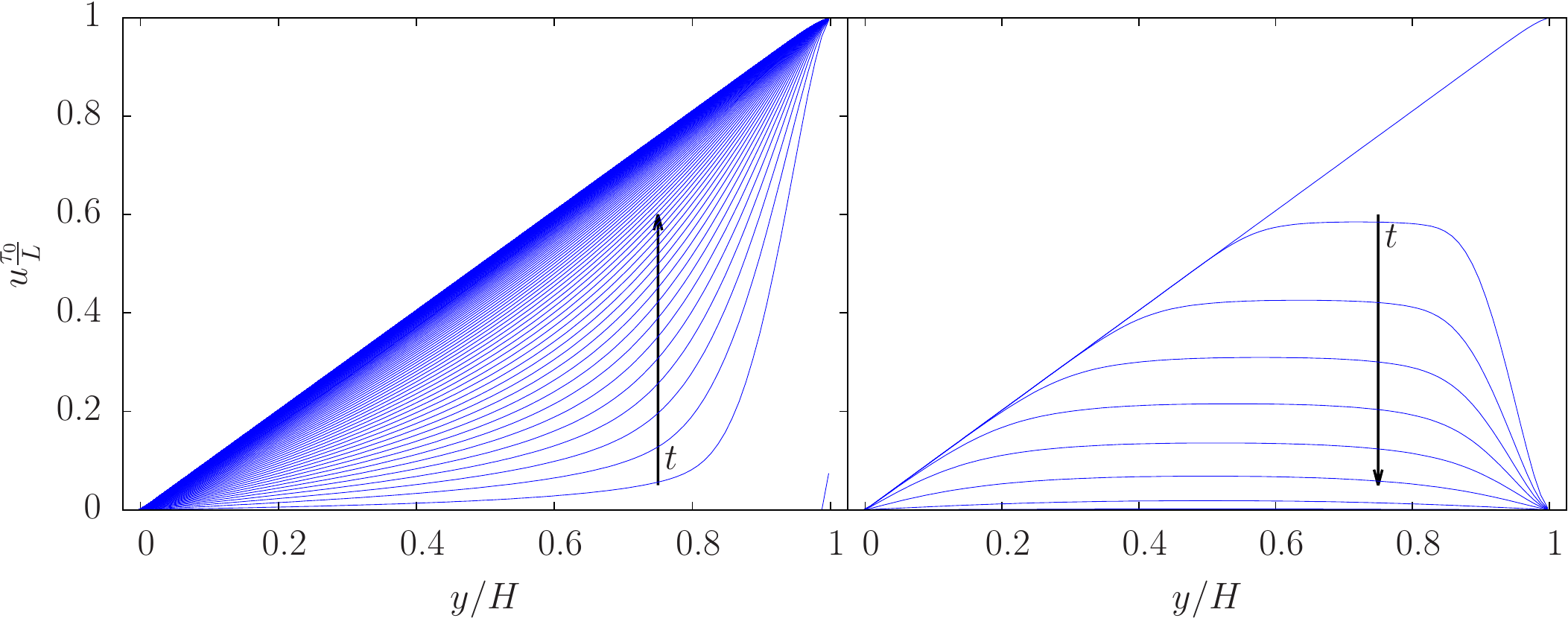}
\caption{\label{stopshear}
  Velocity profiles of a simple shear flow stopped at $t=0$, for various
  times $t>0$.
  Lines are the result of the LB simulation with the generalized Maxwell
  model for $\theta=10^2$, with initial wall velocity $u_0=10^{-2}L/\tau_0$.
}
\end{figure}

The asymmetry between startup and cessation of flow induced by the
non-Newtonian behavior of the fluid is even more clearly seen in the case
of planar Couette flow, where one wall is driven by a fixed velocity that
is instantaneously switched on and off. In Fig.~\ref{stopshear} we
show the resulting build-up and cessation of velocity profiles for this
case. Note that in comparison to the pressure-driven channel, we take the
Couette-flow channel to be of width $H=L$; this accounts for the mirror symmetry
present in the former, but absent in the latter case.

The startup curves shown in Fig.~\ref{stopshear} (left panel) are qualitatively
identical to those of a Newtonian fluid. Note that in the stationary case,
no plug flow develops, since in homogeneous Couette flow the boundary conditions
impose a constant shear rate everywhere. On the other hand, the cessation
curves (right panel) again develop an intermediate plug, starting from the
previously moving wall. They reconfirm qualitatively the results for a
different yield-stress fluid model evaluated within finite-elment
simulations \cite{Chatzimina.2005}.

\section{Conclusions}

We propose a tensorial constitutive equation based on the ideas essential to
nonlinear colloidal rheology, generalizing
a previous model intended to capture certain qualitative
features of a microscopically justified ITT-MCT schematic model proposed
in Ref.~\onlinecite{Brader.2009}. This nonlinear generalization of the
Maxwell model in particular captures the shear-thinning and yield-stress
behavior of glass forming fluids. It is material objective, i.e., its
tensorial structure is compatible with the general laws of continuum
mechanics and coordinate-frame transformations. The model is simple enough
to allow for analytical solutions in certain cases, in particular
pressure-driven Poiseuille flow through a planar channel.

We have developed a modified lattice-Boltzmann simulation scheme to
address the flow of non-Newtonian fluids including the proper tensorial
structure of their constitutive equations. In particular, care has been
taken to inclue non-Newtonian flow-induced pressure differences that
arise through normal-stress differences. These pressure differences cause
the pressure to rise in the plug at the center of the channel. This pressure
variation couples, at the LB level, to a density variation (not unlike
the hydrodynamic interaction effect discussed by Nott and Brady
\cite{Nott.1994}). Here it has to be noted that the nonlinear Maxwell model
we study exhibits rather large normal-stress differences, exaggerating
this effect. This may explain why in MD simulations of plug flow of
glass-forming fluids \cite{Varnik.2008} find only a minor cross-channel
variation in density.

At present we restrict
ourselves to a confirmation of the analytical result in planar flow, where
these normal stresses do not couple back to the flow field. But our method
is easily applied to cases where flow-density couplings become important
and may give rise to nontrivial shear localization \cite{Besseling.2010}.
While the Maxwell relaxation time $\tau$ has been kept constant above,
it should, by its microscopic physical motivation, sensitively depend
on the local density. We leave this extension for further studies.

The modified LB simulation scheme proposed here is found to give accurate
results in the steady-state flow for the shear-thinning model involving
a spread in relaxation times as large as a factor $10^3$. Larger
differences could probably be handled, but at the cost of much finer
grid resolutions and hence computing time. The study of startup flow
and cessation of flow demonstrate that also beyond the steady state,
the proposed modification of the LB algorithm gives accurate results.
In particular, it is capable of reproducing the finite stopping-time
singularity that is typical of yield-stress fluids. Here, the finite
relaxation time $\tau$ introduced in the Maxwell model serves as a natural
regularization parameter.

Other numerical schemes frequently used to simulate non-Newtonian fluid
flows include finite-element and finite-volume modeling, either coupled with
constitutive-equation solvers (see, e.g.,
Refs.~\onlinecite{Malkus.1985,Saramito.1995,Dean.2007}), or for the simpler
generalized-Newtonian fluids \cite{Hron.2000}.
Also within these schemes, the inclusion of non-trivial constitutive
equations poses subtleties. The LB method is a viable alternative that
is computationally efficient.

\begin{acknowledgments}
We thank M.~E.~Cates and D.~Marenduzzo for valuable discussions, and
M.~Fuchs for many fruitful discussions and encouragement.
Th.~V.\ thanks for funding through the Helmholtz-Gemeinschaft,
Young Investigator Group HGF VH-NG 406, and DFG Research Unit FOR1394
``Nonlinear Response to Probe Vitrification'', project P3.
S.~P.\ and Th.~V.\ thank for funding and travel support through
the Zukunftskolleg, Universit\"at Konstanz.

\end{acknowledgments}

\bibliography{lit}
\bibliographystyle{apsrev4-1}

\end{document}